# Electrical Seebeck-Contrast Observation of Magnon Hall Effect in Topological Ferromagnet $Lu_2V_2O_7$/Heavy Metal Heterostructures

Jinsong Xu[1*], Jiaming He[2], J.-S. Zhou[2], Danru Qu[3], Ssu-Yen Huang[4], and C. L. Chien[1,4*]

[1] *Department of Physics and Astronomy, Johns Hopkins University, Baltimore, Maryland 21218, USA*

[2] *Department of Mechanical Engineering, University of Texas at Austin, Austin, Texas 78712, USA*

[3]*Center for Condensed Matter Sciences, National Taiwan University, Taipei 10617, Taiwan*

[4]*Department of Physics, National Taiwan University, Taipei 10617, Taiwan*

*Corresponding author. Email: jxu94@jhu.edu (J.X.); clchien@jhu.edu (C.L.C.)

**Abstract**

**The observation of the magnon Hall effect (MHE) has relied solely on the challenging measurement of the thermal Hall conductivity. Here, we report a highly sensitive electrical Seebeck-contrast method for the observation of MHE in $Lu_2V_2O_7$/heavy metal heterostructures, that is highly desirable for the exploration of new MHE materials and their applications. Using measuring wires with very different Seebeck coefficients, we established a general method that can separate contributions (e.g., MHE) that generates a lateral temperature drop, from those [e.g., anomalous Nernst effect (ANE) and spin Seebeck effect (SSE)] that generate a lateral electric field. We show that a suitable heavy metal overlayer can eliminate the inherent ANE and SSE signals from the semiconducting $Lu_2V_2O_7$. The MHE in $Lu_2V_2O_7$ is quasi-isotropic among crystals with different orientations. In addition to the previously reported transverse MHE under an in-plane temperature gradient, we have uncovered longitudinal MHE under an out-of-plane temperature gradient.**

## I. INTRODUCTION

Magnons in insulating magnets have the unique attributes of delivering spin angular momentum coherently over long distances and without Joule heat dissipation [1]. The excitation, propagation, control, and detection of magnons have been the central topics in the field of magnonics. Recently, the discovery of the magnon Hall effect (MHE), an analogue of the anomalous Hall effect (AHE), in topological ferromagnet $Lu_2V_2O_7$, has attracted much attention [2–14], offering opportunities for exploring topologically protected magnon transport [5,6,15–17].

In a ferromagnetic (FM) material, the Berry curvature due to the spin-orbit interaction gives rise to the AHE [18]. The Berry curvature can also act on other particles and even charge neutral entities, such as magnons, as realized in the MHE, which in turn provides insight into the magnon band topology. In the MHE, the magnon flow driven by the temperature gradient with the spin index set by the magnetization direction, generates a temperature difference in the third direction, resulting from the magnon flow deflected by the Berry curvature. However, the observation of MHE has thus far relied only on the challenging thermal Hall conductivity measurements [4,7–9,13,14], thus impeding general applications. It is also challenging to detect the predicted topological magnon surface states due to the limited experimental probes [15,17,19], let alone the utilization of these surface states. While one can inject and detect magnons in magnetic insulators electrically via a heavy metal (HM) overlayer [20,21], the HM itself may also modify the magnon transport [22,23]. Therefore, it remains unclear whether electrical detection of MHE is even feasible.

The MHE differs from the other related Hall effects [e.g., the ordinary Hall effect (OHE), the anomalous Hall effect (AHE), and inverse spin Hall effect (ISHE)] in one important aspect. The OHE, the AHE [and the related ANE], and the ISHE in conjunction with the SSE are electric field ($E$) effects. For example, in OHE, $E_{OHE} \propto j_c \times B$, in AHE, $E_{AHE} \propto j_c \times M$, where $j_c$ is the charge current, $B$ and $M$ are the magnetic field and magnetization respectively, and in ISHE, $E_{ISHE} \propto j_s \times \sigma$, where $j_s$ is the spin current and $\sigma$ is the spin index. In these cases, one detects a lateral electric voltage due to the specific electric field. In contrast, in the MHE, the deflected magnon flow due to the Berry curvature, produces a temperature gradient and detected as a temperature drop. This key difference allows us to exploit the Seebeck-contrast method to differentiate and extract the MHE in the electrical measurements.

In this work, we establish an electrical Seebeck-contrast technique for the observation of MHE in $Lu_2V_2O_7$/HM heterostructures, where the HM is Pt or W, under an in-plane temperature gradient. We exploit the Seebeck contrast of different metals to separate the MHE and the ANE/SSE contributions. In $Lu_2V_2O_7$/HM with different HMs and thicknesses, we find that the MHE is independent of the HM overlayer, and the ANE/SSE contribution within $Lu_2V_2O_7$ can be eliminated by a thick HM overlayer. The ease of the electrical detection of MHE also allows us to uncover longitudinal-MHE in $Lu_2V_2O_7$/HM heterostructures under an out-of-plane temperature gradient, which should also be applicable in thin films of MHE materials. The electrical Seebeck-contrast method, which provides fast screening of MHE candidate materials, is also a general method that separates contributions that generate a lateral electric field (e.g., OHE, AHE, ANE, and ISHE) from those that generate a lateral temperature gradient (e.g., MHE).

The longitudinal MHE in $Lu_2V_2O_7$/HM reported here, and the well-established longitudinal SSE in YIG/HM, share the same measuring scheme. Our established protocol provides clear differentiation between MHE and SSE, and can be exploited to establish thermal spin transport in other magnetic insulators. The electrical detection of MHE paves the way for controlling spin thermal transport based on the thermal Hall effect, and possibility of electrical detection and utilization of topological magnon surface states.

## II. EXPERIMENTAL DETAILS

Single crystals $Lu_2V_2O_7$ were grown by the floating zone method with an image furnace (NEC SC-M35HD). A rod was prepared from thoroughly mixed stoichiometric amount of $Lu_2O_3$ (Alfa Aesar, 99.99%) and $V_2O_4$ (Alfa Aesar, 99.99%); the resulting rod was then annealed inside the image furnace with Ar gas environment. The crystal growth was carried out in the same gas environment with a growing speed of 5 mm/h. The $Lu_2V_2O_7$ ingot is black in color with a shiny surface, and the crystal quality was verified by Laue X-ray back reflection. The refinement of powder X-ray diffraction pattern with the cubic pyrochlore structural model and space group *Fd-3m* (No. 227) gives the lattice parameters of *a* =9.936 (1) Å. All disks of oriented crystals were polished with $Al_2O_3$ lapping papers ranging from 30 μm to 0.3 μm. Heavy metal W and Pt layers were deposited on the polished crystal disks by DC magnetron sputtering at room temperature in a vacuum chamber with a base pressure of $1.0 \times 10^{-8}$ Torr.

The Lu$_2$V$_2$O$_7$/HM devices were thermally anchored to two Cu blocks via silicone thermal pads and one of the Cu blocks was heated by a resistive heater. The temperature of two Cu blocks ($T_{hot}$ and $T_{cold}$) was monitored by two Cernox temperature sensors, which gives the temperature difference $\Delta T \equiv T_{hot} - T_{cold}$, and the sample temperature $T_{avg} \equiv (T_{hot} + T_{cold})/2$. The temperature gradient $\nabla T$ is calculated via $\nabla T \equiv \Delta T / L_T$, where $L_T$ is the distance between two temperature sensors. The value $|\nabla T|$ used in this study is on the order of 4 K/mm. The magneto-thermal voltage $V$ was measured with a Keithley 2182 nanovoltmeter.

## III. RESULTS AND DISCUSSION

### 1. Electrical and magnetic properties of Lu$_2$V$_2$O$_7$ and Lu$_2$V$_2$O$_7$/HW heterostructures

Lu$_2$V$_2$O$_7$ is a semiconducting ferromagnet with the pyrochlore structure [4,7,24]. The Dzyaloshinskii-Moriya interaction (DMI) resulting from the broken inversion symmetry gives rise to the MHE [4,7]. We used crystals with [100], [111] or [110] direction oriented normal to the surface. We primarily used the [100]-oriented Lu$_2$V$_2$O$_7$ crystals to establish the electrical observation of MHE unless otherwise noted.

The magnetic properties of Lu$_2$V$_2$O$_7$ single crystals have been characterized. The spontaneous magnetization $M$ emerges below the Curie temperature of $T_c$ = 70 K (Fig. S1 [25]) and saturates at a relatively low magnetic field (less than 5 kOe), similar to those previously reported [4,26]. The saturated $M$ is close to the spin-only moment of 1 $\mu_B$ per V$^{4+}$ ion.

The electrical resistivity of Lu$_2$V$_2$O$_7$ is highly temperature dependent and increases rapidly with decreasing temperature (Fig. 1(a)), indicating a semiconducting behavior with especially large resistivities below $T_c$. Fig. 1(b) shows the field dependence of the anomalous Hall resistivity $\rho_{AHE}$ at different temperatures, where the linear background from the ordinary Hall effect has been subtracted. Because AHE is antisymmetric with respect to the magnetic field $H$, the $H$-symmetric component has also been subtracted. The same procedure has also been applied to MHE data [4,7,25]. Similar to the magnetization results, AHE is clearly present below $T_c$ = 70 K and saturates under a magnetic field of less than 5 kOe. The temperature dependent $\rho_{AHE}$ at $H$ = 5 kOe is shown in Fig. 1(c), where $\rho_{AHE}$ reaches a maximum value at 50 K, then decreases at lower temperatures, and changes sign at 30 K. The sign change is also observed in the ANE (Fig. S3 [25]). This strong temperature dependence of AHE and ANE are likely the consequences of strong Berry curvature effects in $Lu_2V_2O_7$ [18,27–29].

We note the AHE and ANE results at 30 K in Fig. 1(b) and Fig. S3 [25] are considerably noisier because of the much higher resistivity, which also impedes electrical detection of MHE in $Lu_2V_2O_7$. However, after depositing a HM overlayer, the resistivity of the $Lu_2V_2O_7$/HM and the electrical noise (Fig. S4 [25]) are much reduced, enabling the electrical detection of MHE. As shown in Fig. 1(d) and 1(e), $\rho_{AHE}$ of $Lu_2V_2O_7$/HM with 3 nm Pt and W overlayer, respectively, are three orders of magnitude smaller than that in $Lu_2V_2O_7$ because of the shunting effect. The AHE eventually diminishes to an undetectable level with 15 nm Pt (Fig. 1(f)), which also suppresses the ANE/SSE contribution in the electrical detection of MHE, as we discuss later.

## 2. Seebeck-contrast technique: electrical observation of MHE in $Lu_2V_2O_7$/HM heterostructures

The observation of MHE has thus far relied only on the challenging spin-dependent thermal Hall conductivity measurements [4,7–9,13,14], where two temperature sensors are attached to opposite sides of the sample for measuring the induced temperature difference $\Delta T_y$ as a result of the applied temperature gradient $\nabla_x T$. The thermal Hall conductivity measurement is time-consuming and involves complex instrumentations. Here, we provide a more straightforward and sensitive electrical Seebeck-contrast method to detect MHE, that provides fast screening of MHE candidate materials and is highly desirable for the exploration and applications of MHE. Furthermore, we experimentally demonstrate that MHE shares the same symmetry as those of ANE and SSE, thus are entangled and cannot be distinguished in a single measurement. However, the Seebeck-contrast technique can unequivocally distinguish MHE from ANE/SSE, by performing measurements using two different wires with highly contrasting Seebeck coefficients. This technique is useful for establishing SSE or MHE in new materials.

The set-up for the electrical detection of MHE is shown in Fig. 2(d). Under an applied $\nabla_x T$ and $H_z$, the large Berry curvature in $Lu_2V_2O_7$ gives rise to MHE, where the deflected magnons flow causes a spin-dependent temperature difference $\Delta T_y$ [4,7]. The induced $\Delta T_y$ is converted into a thermal voltage via the thermocouple effect due to the different Seebeck coefficients of the HM and probing wire,

$$V_{\text{HM}}^{\text{wire}}(\text{MHE}) = (S_{\text{wire}} - S_{\text{HM}})\Delta T_y, \quad (1)$$

where the wire is made of Cu or $Cu_{55}Ni_{45}$ (CuNi for short). Because the rectangular-shaped crystals have different lengths along the crystal edges, for direct comparison among different measurements, we use normalized voltage $\alpha$, normalized by the respective sample dimension, i.e., $\alpha = \frac{V/L_V}{\Delta T_x/L_T}$, where $V$ is the measured voltage, $L_V$ and $L_T$ are the sample lengths between the two voltage leads and the two temperatures.

The MHE induced $\Delta T$ is proportional to $\nabla T \times M$, therefore the MHE induced voltage follows $\nabla V(\text{MHE}) \propto \nabla T \times M$, which has the same symmetry as those of the two other well-known spin caloritronic effects of ANE and SSE. For ANE in a conducting ferromagnet or insulating ferromagnet/HM bilayer due to magnetic proximity effect, the transverse voltage also fulfills $\nabla V(\text{ANE}) \propto \nabla T \times M$. For both longitudinal SSE [30] and recently discovered unconventional transverse SSE [31,32], $\nabla T$ generates a spin current $J_s$ with polarization $\sigma$, and detected as a voltage via the ISHE, following $\nabla V(\text{SSE}) \propto J_s \times \sigma \propto \nabla T \times M$. Therefore, the total voltage measured using the set-up in Fig. 2(d) is

$$V_{\text{HM}}^{\text{wire}} = V_{\text{HM}}^{\text{wire}}(\text{MHE}) + V_{\text{HM}}^{\text{wire}}(\text{ANE/SSE}), \quad (2)$$

i.e., the measured voltage contains contributions from both MHE and ANE/SSE.

Since MHE, ANE and SSE all have the same symmetry, they cannot be separated in a single measurement. However, the difference between MHE and ANE/SSE can be unequivocally distinguished by two measurements using two wires with very different Seebeck coefficients [10,33]. The ANE and SSE signal of $V_{\text{HM}}^{\text{wire}}(\text{ANE/SSE})$ would be the same when measured using Cu and CuNi wires, whereas the MHE signal of $V_{\text{HM}}^{\text{wire}}(\text{MHE})$ would be very different because of the large difference between $S_{\text{Cu}}$ and $S_{\text{CuNi}}$. From Eq. (1) and (2), the voltage *difference* between the two measurements using Cu and CuNi wires contains *only* the MHE part with the ANE/SSE parts removed,

$$\Delta V_{\text{HM}} \equiv V_{\text{HM}}^{\text{Cu}} - V_{\text{HM}}^{\text{CuNi}} = V_{\text{HM}}^{\text{Cu}}(\text{MHE}) - V_{\text{HM}}^{\text{CuNi}}(\text{MHE}) = (S_{\text{Cu}} - S_{\text{CuNi}})\Delta T_y, \quad (3)$$

As shown in Fig. 2(a), in Lu$_2$V$_2$O$_7$ without the HM overlayer, the voltage $\alpha^{Cu}$ measured using Cu wires and the voltage $\alpha^{CuNi}$ measured using CuNi wires are essentially the same, with negligible difference $\Delta\alpha$ well within the noise. This indicates that the weaker MHE contribution is buried within the noise (on the order of 100 nV/K) of the dominating ANE contribution, largely due to the high resistivity of Lu$_2$V$_2$O$_7$ at low temperatures. However, after depositing a 3 nm Pt layer, in Lu$_2$V$_2$O$_7$/Pt(3 nm), the value of $\alpha_{Pt}^{Cu}$ is distinctively different from that of $\alpha_{Pt}^{CuNi}$, in fact with the opposite sign, as shown in Fig. 2(b). The difference $\Delta\alpha_{Pt} = \alpha_{Pt}^{Cu} - \alpha_{Pt}^{CuNi} = 8.68 \pm 0.46$ nV/K. These results clearly show the otherwise small MHE in bare Lu$_2$V$_2$O$_7$ can be extracted using Lu$_2$V$_2$O$_7$/Pt heterostructure via the Seebeck-contrast technique exploiting the HM overlayer.

The results of Lu$_2$V$_2$O$_7$/W(3 nm) are shown in Fig. 2(c), where the red curve of $\alpha_W^{Cu}$ and the blue curve of $\alpha_W^{CuNi}$ are very different from the red curve of $\alpha_{Pt}^{Cu}$ and the blue curve of $\alpha_{Pt}^{CuNi}$ in Fig. 2(b). However, the black curve of $\Delta\alpha_W = \alpha_W^{Cu} - \alpha_W^{CuNi}$ in Fig. 2(c) is the *same* as the black curve of $\Delta\alpha_{Pt} = \alpha_{Pt}^{Cu} - \alpha_{Pt}^{CuNi} = 8.68 \pm 0.46$ nV/K in Fig. 2(b). The black curves in Fig. 2(b) with 3 nm Pt and in Fig.2(c) with 3 nm W are the MHE intrinsic to Lu$_2$V$_2$O$_7$, irrespective of the HM overlayers. These examples illustrate clearly the new method of exploiting Seebeck-contrast measurements with HM overlayer to extract the weaker MHE signals from Lu$_2$V$_2$O$_7$. The results at different temperatures are shown in Fig. S5 [25].

We further explore the dependence of MHE on $\nabla_x T$. Fig. 2(e) shows the MHE voltage $\Delta V_{Pt}$ at three different temperature gradients at $T_{avg} = 60$ K. $\Delta V_{Pt}$ decreases when reducing $\Delta T_x$ from 40 K to 20 K, and changes sign when $\Delta T_x$ of -16 K is applied. The MHE signal depends linearly on $\Delta T_x$ as shown in Fig. 2(f), hence linear MHE in Lu$_2$V$_2$O$_7$.

## 3. Separating MHE and ANE/SSE contributions in Lu$_2$V$_2$O$_7$/HM heterostructures

The thermal transport measurements in Lu$_2$V$_2$O$_7$ contain contributions from the dominant ANE/SSE and the weaker MHE. The Seebeck-contrast technique can identify the MHE part and can also extract the ANE/SSE contributions.

From Eq. (3), using the known values of $S_{Cu} - S_{CuNi}$, from the measured $\Delta V_{HM}$, we obtain $\Delta T_y$ induced from MHE. Substituting it back into Eq. (1) we obtain the MHE contribution of

$$V_{HM}^{wire}(MHE) = (S_{wire} - S_{HM})\Delta T_y = \frac{S_{wire}-S_{HM}}{S_{Cu}-S_{CuNi}} \Delta V_{HM}, \qquad (4)$$

Together with Eq. (2), the ANE/SSE contribution is

$$V_{HM}^{wire}(ANE/SSE) = V_{HM}^{wire} - V_{HM}^{wire}(MHE) = V_{HM}^{wire} - \frac{S_{wire}-S_{HM}}{S_{Cu}-S_{CuNi}} \Delta V_{HM}, \qquad (5)$$

Therefore, as shown in Eq. (4) and (5), the Seebeck-contrast technique can unequivocally distinguish MHE from ANE/SSE, crucial for establishing SSE or MHE in new materials.

As shown in Fig. 2, there is a single value of various $\alpha$ values at the saturation value (e.g., at $H_z$ = 3 kOe). The saturation values of $\alpha_{Pt}^{Cu}$ and $\alpha_{Pt}^{CuNi}$ at 3 nm Pt in Fig. 2(b) and the values of $\alpha_W^{Cu}$ and $\alpha_W^{CuNi}$ at 3 nm W in Fig. 2(c) at 60 K are shown in Fig. 3(a), 3(b) with circles, and the $\Delta\alpha_W = \Delta\alpha_{Pt}$ value as a circle in Fig. 3(c). These temperature dependence results show that different HM overlayers impact the individual measured voltage $\alpha_{HM}^{wire} = \alpha_{HM}^{wire}(MHE) + \alpha_{HM}^{wire}(ANE/SSE)$, because not only $\alpha_{HM}^{wire}(MHE)$ changes due to the different Seebeck coefficients of the HM layers, but also $\alpha_{HM}^{wire}(ANE/SSE)$ varies due to the different shunting effect and/or different spin Hall angles of the HM. However, as shown in Fig. 3(c), the value of $\Delta\alpha = \alpha_{HM}^{Cu} - \alpha_{HM}^{CuNi}$ is *independent* of the HM nor its thickness, because it arises from MHE and dictated by the Berry curvature in Lu$_2$V$_2$O$_7$, and the Seebeck coefficients of the probing wires. Given $S_{Cu}$ and $S_{CuNi}$ [34,35], the estimated $\nabla T_y/\nabla T_x$ is on the order of $8 \times 10^{-4}$. This is comparable but slightly smaller than the previous results $1.0 \times 10^{-3}$ measured at 60 K by thermal Hall conductivity method [4,7]. In the calculation, we have used $\nabla_x T = \Delta T_x/L_T$. However, in reality, there is a temperature drop between thermal baths and the sample [36], which we can consider as an effective length $L_C$ in addition to the actual sample length $L_T$, and the actual temperature gradient is $(\nabla_x T)_{act} = \Delta T_x/(L_T + L_C)$. Therefore, $\frac{\nabla T_y}{\nabla T_x} = (\frac{\nabla T_y}{\nabla T_x})_{act} \cdot \frac{L_T}{L_T+L_C}$. The estimated $\frac{\nabla T_y}{\nabla T_x}$ is the lower bound of the actual value $(\frac{\nabla T_y}{\nabla T_x})_{act}$. Nevertheless, these results demonstrate the validity of the electrical detection method and its effectiveness in measuring MHE in Lu$_2$V$_2$O$_7$.

Based on the known values of $S_W$ and $S_{Pt}$ [34,37], we can separate MHE and ANE/SSE contributions using Eq. (4) and Eq. (5). The results using Cu wire $\alpha_{HM}^{Cu}(MHE)$ and $\alpha_{HM}^{Cu}(ANE/SSE)$ are summarized in Fig. 3(d) and 3(e). Comparing samples with 3 nm Pt and 3 nm W, $\alpha_{HM}^{Cu}(MHE)$ is very different from each other because it depends on the different Seebeck coefficients of Pt and W, while $\alpha_{HM}^{Cu}$(ANE/SSE) is similar to each other as it is mainly determined by the resistivity of Pt and W due to the shunting effect. This is further confirmed by comparing samples with 3 nm and 15 nm Pt, where $\alpha_{HM}^{Cu}$(MHE) is the same assuming the Seebeck coefficient of Pt is unaffected by its thickness, and $\alpha_{HM}^{Cu}$(ANE/SSE) diminishes with 15 nm Pt because of the shunting effect. Here, 15 nm Pt is sufficient to suppress the ANE/SSE contribution in $Lu_2V_2O_7$. Furthermore, the magnitude of MHE and ANE/SSE can be comparable, of either sign depending on the HM. This means, for the SSE study, one can obtain completely misleading results when one fails to include possible MHE contribution. It is also noteworthy that, AHE and ANE are also closely related. Since $\rho_{AHE}$ of $Lu_2V_2O_7$/3nm W is larger than $\rho_{AHE}$ of $Lu_2V_2O_7$/3nm Pt due to the larger resistance of W, one would expect larger $\alpha_W^{Cu}$(ANE) than $\alpha_{Pt}^{Cu}$(ANE), which is opposite to our observation in Fig. 3(e). Therefore, there is indeed some contribution from the opposite spin Hall angle of W and Pt, like unconventional SSE. Further investigations are required to explore the possible role of topological magnon surface states and magnon Weyl points [5,6,15,17]. Similar behaviors have been observed with the CuNi probing wire, with results shown in Fig. 3(f) and 3(g).

## 4. Transverse and longitudinal MHE in $Lu_2V_2O_7$/HM heterostructures

Although there are some theoretical predictions on the dependence of MHE on the field direction and crystal orientation, a detailed experimental investigation remains scarce [4,38–40]. Fig. 4(a) shows the results of MHE signal $\Delta\alpha_{Pt}$ as a function of $H_z$ under $\nabla_x T$. However, as shown in Fig. S6(a) [25] when the magnetic field is applied along x-direction under $\nabla_x T$, both MHE and ANE/SSE signal vanish because of the orthogonal relationship $\nabla V \propto \nabla T \times M$, confirming the MHE and ANE/SSE observed in Fig. 2 arise from in-plane temperature gradient $\nabla_x T$ and out-of-plane magnetization $M_z$. Furthermore, [100], [111] and [110]-oriented $Lu_2V_2O_7$ crystals show similar results and with similar temperature dependence as shown in Fig. 4(b). We observe no discernible difference among these crystals, indicating that the MHE in $Lu_2V_2O_7$ is essentially isotropic. These results are in good agreement with those previously measured by thermal Hall conductivity [4].

The MHE in $Lu_2V_2O_7$ to date has only been realized under an in-plane temperature gradient (e.g., $\nabla_x T$ in Fig. 2(d)), the so-called transverse injection scheme, and detected by thermal Hall conductivity (previous work) [4,7] or the electrical measurement (this work), along a direction also in-plane but perpendicular to the injection direction (e.g., $V_y$ in Fig. 2(d)). In contrast, the SSE in YIG can only be realized in the so-called longitudinal scheme under an out-of-plane temperature gradient but not the transverse scheme [41–44]. In this work, with the more straightforward and sensitive electrical detection of MHE at hand, we have further explored the MHE under an out-of-plane temperature gradient $\nabla_z T$. There is indeed MHE signal $\Delta\alpha_{Pt}$ when the magnetic field is along the *x*-direction (Fig. 4(c) and 4(d)), but not along the *z*-direction (Fig. S6(b) [25]), which further confirms the aforementioned orthogonal relationship of $\nabla V \propto \nabla T \times M$. However, comparing with the $\nabla_x T$ case, the MHE signal $\Delta\alpha_{Pt}$ is one order of magnitude smaller under $\nabla_z T$. This is not contradictory to the isotropic nature of MHE in $Lu_2V_2O_7$ observed in Fig. 4(a) and 4(b), but a reflection of different injection schemes as revealed in the SSE of YIG. The newly observed longitudinal MHE can be understood in terms of the effective length $L_C$ caused by the temperature drop at the thermal contacts, i.e., $\Delta\alpha_{Pt} = (\Delta\alpha_{Pt})_{act} \cdot \frac{L_T}{L_T+L_C}$, where $(\Delta\alpha_{Pt})_{act}$ is the actual value. The sample is a slab with the thickness (~1 mm) much smaller than the lateral length (~10 mm), therefore $L_T$ under $\nabla_z T$ is much smaller than that under $\nabla_x T$, resulting in a smaller calculated $\Delta\alpha_{Pt}$. This is further evidenced by the slightly larger value of $\Delta\alpha_{Pt}$ in the [111]-oriented $Lu_2V_2O_7$ sample because its thickness is 1.5 mm as compared to the 1 mm thickness of [100] and [110]-oriented $Lu_2V_2O_7$. Nonetheless, this is the first observation of longitudinal MHE under $\nabla_z T$ complementing $\nabla V \propto \nabla T \times M$, which has not been previously reported in thermal Hall conductivity measurement. Furthermore, the newly revealed longitudinal MHE should also be applicable to thin films of MHE materials, which would be even more challenging for the conventional thermal Hall conductivity measurement.

## 5. Seebeck-Contrast Electrical Measurement

We have illustrated in detail the Seebeck-contrast electrical measurement in the case of magnon Hall effect in $Lu_2V_2O_7$. This method entails performing voltage measurements using two set of wires made of metals with contrasting Seebeck coefficients. If the voltage measurements using these different wires give the same result, then it is an electrical field effect, as in the cases of ordinary Hall effect, anomalous Hall effect, inverse spin Hall effect, etc. On the other hand, if the voltage measurements yield different results, then it is a temperature gradient effect, as in the case of the magnon Hall effect, which can be revealed by the difference of the two voltages. This Seebeck-contrast electrical method can reveal the nature of various transport phenomena in quantum materials with significant spin-orbit coupling.

**IV. CONCLUSIONS**

In conclusion, we have demonstrated an electrical Seebeck-contrast technique for observing MHE in topological ferromagnet $Lu_2V_2O_7$, simpler than measuring the thermal Hall conductivity and offering a fast screening of MHE candidate materials. We have also discovered longitudinal MHE, in addition to the previously known transverse MHE in $Lu_2V_2O_7$/HM. This new electrical method is applicable to thin films of MHE materials, which would be rather difficult for the conventional thermal Hall conductivity measurement. Because MHE, ANE, and SSE share the same symmetry and may co-exist in the same material as we experimentally demonstrated in $Lu_2V_2O_7$, it is of vital importance to distinguish and identify these different contributions for exploring spin caloritronic phenomena. In particular, the established Seebeck-contrast technique using two probing wires of different Seebeck coefficients can separate MHE, which generates a lateral temperature difference, from ANE/SSE in $Lu_2V_2O_7$/HM, that generates a lateral electric field and a voltage drop. This protocol can be exploited to identify spin current effect in quantum materials.

## ACKNOWLEDGEMENTS

This work was supported by NSF DMREF Awards No. 1729555, No. 1949701, and partially NSF MRSEC DMR-2308817. D.Q. and S.Y.H. were supported by the National Science and Technology Council under Grants No. NSTC 110–2112-M-002–047-MY3 and No.NSTC 112–2123-M-002–001, respectively. J.X. has been partially supported by DOE Basic Energy Science Award No. DE-SC0009390.

## COMPETING INTERESTS

The authors declare no competing interests.

## DATA AVAILABILITY

The data that support the findings of this study are available from the corresponding author upon reasonable request.

## *CORRESPONDING AUTHORS

jxu94@jhu.edu (J.X.); clchien@jhu.edu (C.L.C.)

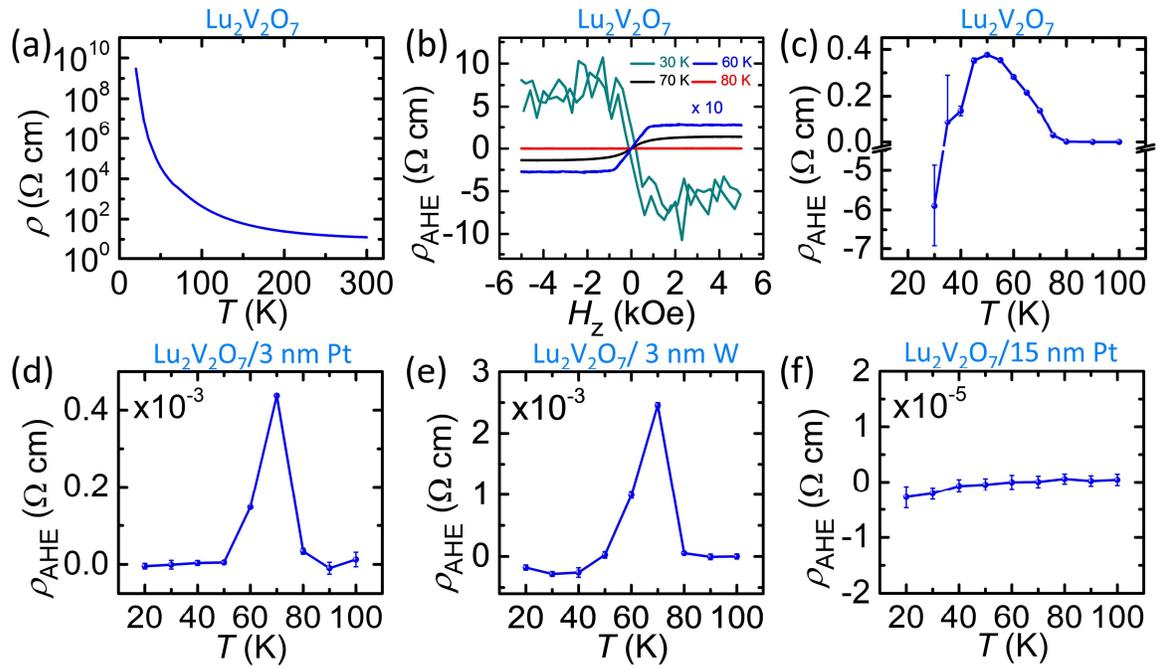

**FIG. 1.** (a) Temperature dependence of $Lu_2V_2O_7$ resistivity. (b) AHE in $Lu_2V_2O_7$ at different temperatures. (c)-(f) Temperature dependence of $\rho_{AHE}$ in $Lu_2V_2O_7$, $Lu_2V_2O_7$/3 nm Pt, $Lu_2V_2O_7$/3 nm W and $Lu_2V_2O_7$/15 nm Pt respectively.

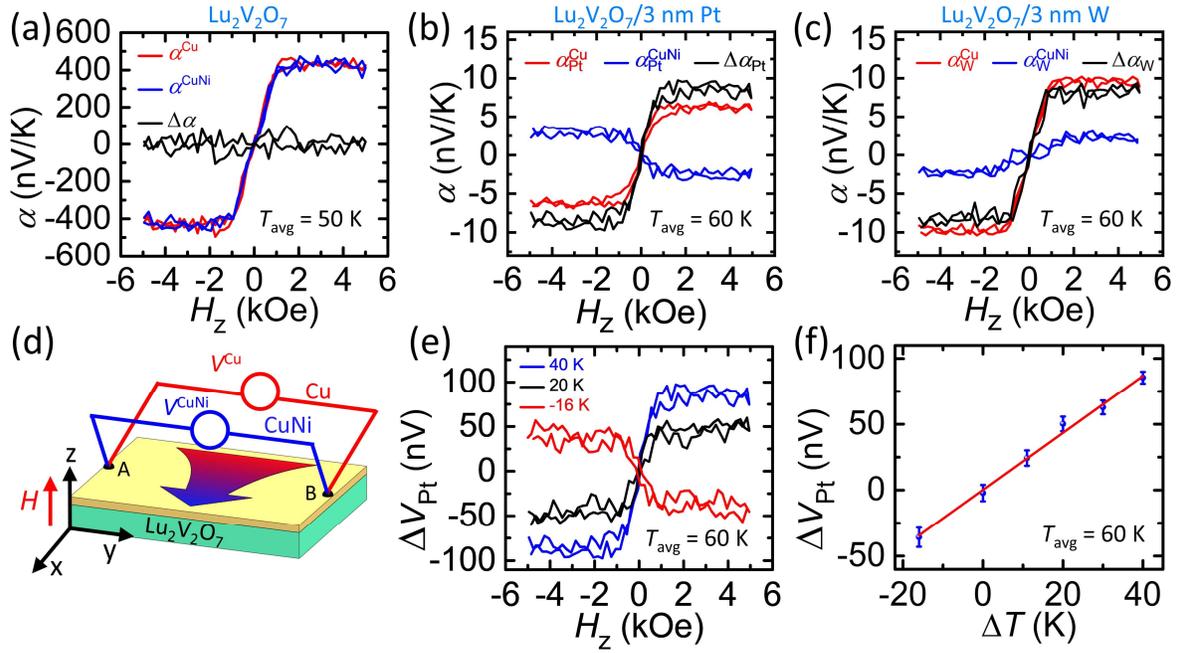

**FIG. 2.** (a)-(c) Voltages as a function of applied magnetic field, measured using Cu wire (red curve), CuNi wire (blue curve) and the difference between them (black curve) in $Lu_2V_2O_7$, $Lu_2V_2O_7$/3 nm Pt, and $Lu_2V_2O_7$/3 nm W respectively at specific temperature. (d) The schematic for electrical detection of MHE. (e) MHE as a function of applied magnetic field at different temperature gradients in $Lu_2V_2O_7$/3 nm Pt. (f) Temperature gradient dependence of MHE in $Lu_2V_2O_7$/3 nm Pt.

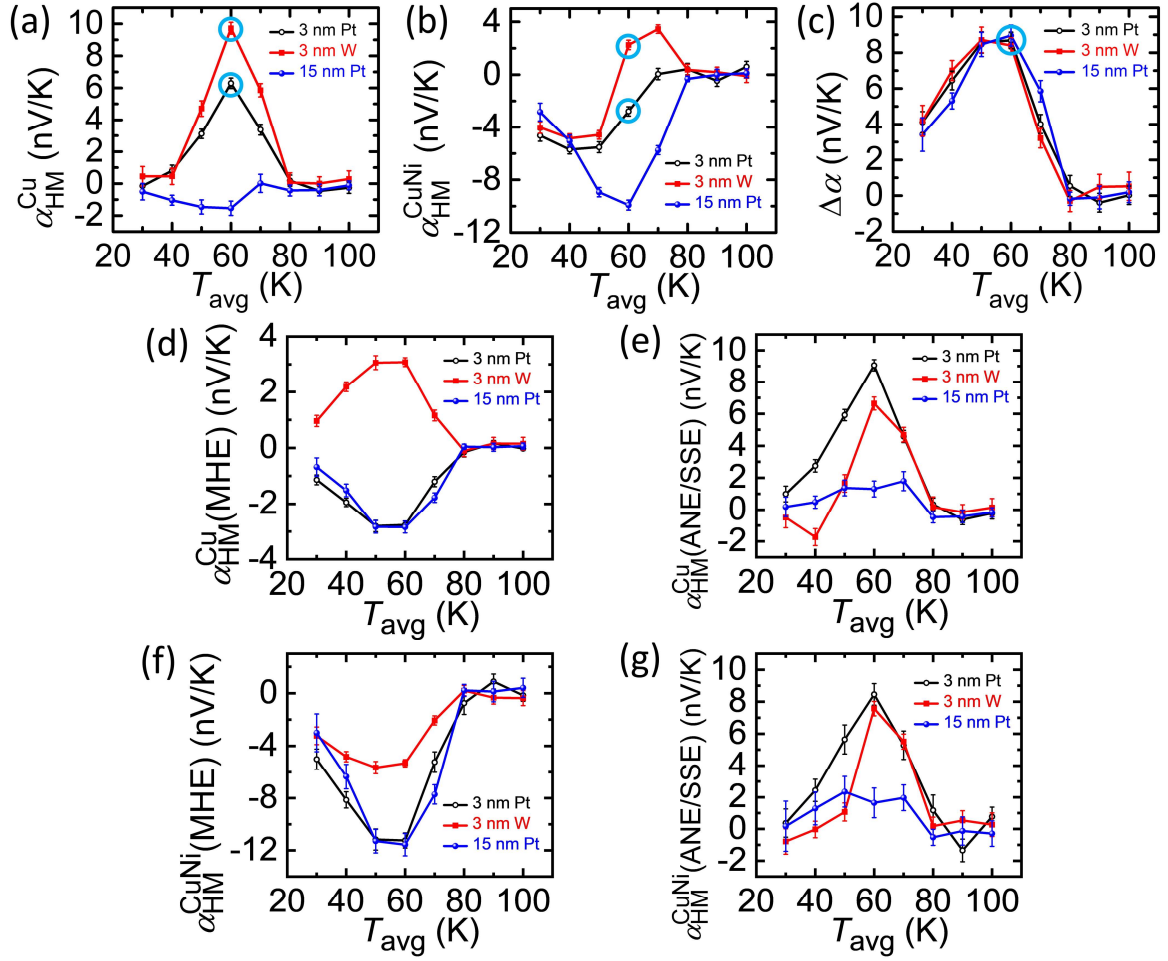

**FIG. 3.** (a)-(c) Temperature dependence of $\alpha_{HM}^{Cu}$, $\alpha_{HM}^{CuNi}$ and $\Delta\alpha = \alpha_{HM}^{Cu} - \alpha_{HM}^{CuNi}$ in $Lu_2V_2O_7$ with 3 nm Pt, 3 nm W and 15 nm Pt overlayers. (d)-(e) Temperature dependence of MHE contribution $\alpha_{HM}^{Cu}$ (MHE) and ANE/SSE contribution $\alpha_{HM}^{Cu}$ (ANE/SSE) in $Lu_2V_2O_7$ with 3 nm Pt, 3 nm W and 15 nm Pt overlayers, measured using Cu wire. (f)-(g) Temperature dependence of MHE contribution $\alpha_{HM}^{CuNi}$ (MHE) and ANE/SSE contribution $\alpha_{HM}^{CuNi}$ (ANE/SSE) in $Lu_2V_2O_7$ with 3 nm Pt, 3 nm W and 15 nm Pt overlayers, measured using CuNi wire.

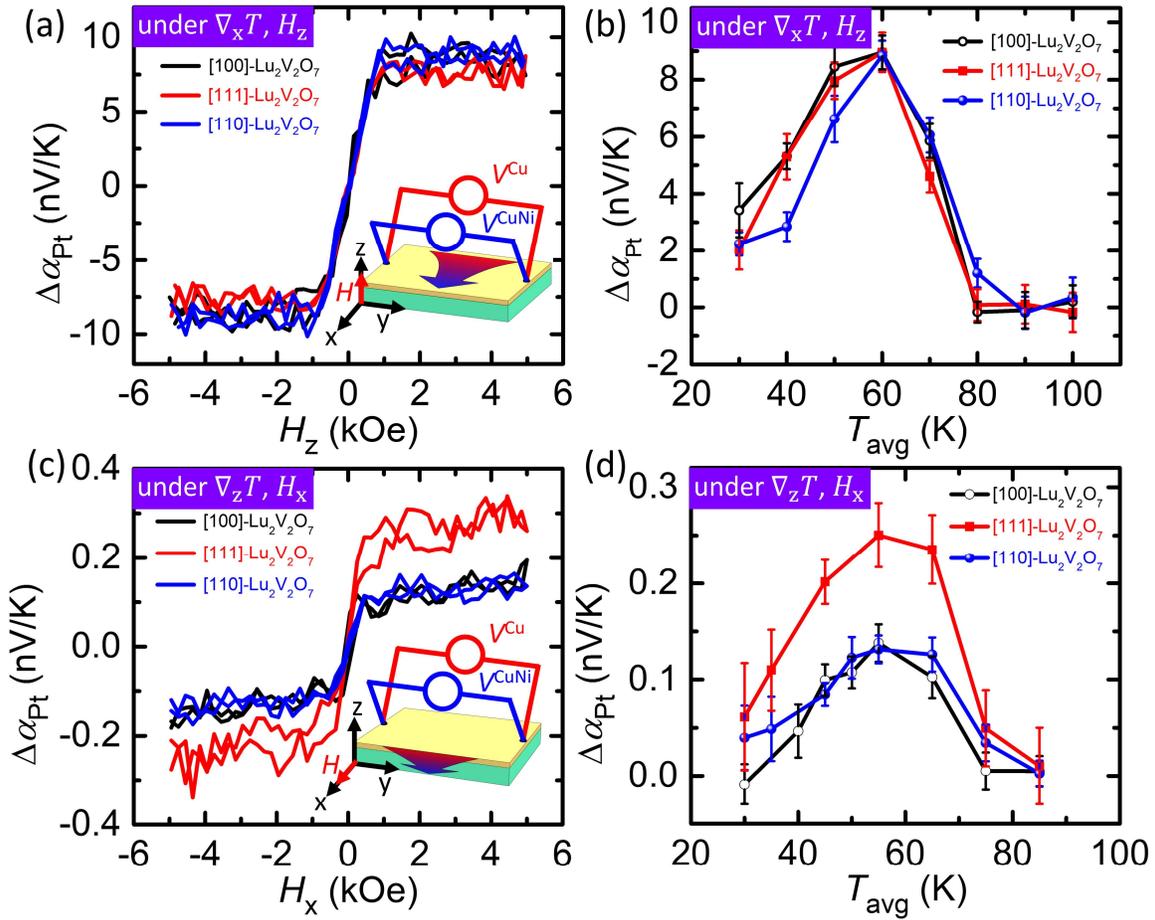

**FIG. 4.** (a) Transverse MHE along different crystal orientations under $\nabla_x T$ and $H_z$ at $T_{avg}$ = 60 K. (b) Temperature dependence of transverse MHE along different crystal orientations under $\nabla_x T$ and $H_z$. (c) Longitudinal MHE along different crystal orientations under $\nabla_z T$ and $H_x$ at $T_{avg}$ = 55 K. (d) Temperature dependence of longitudinal MHE along different crystal orientations under $\nabla_z T$ and $H_x$. Black, red and blue curves represent [100], [111] and [110]-oriented $Lu_2V_2O_7$, respectively. The inset in (a) and (c) are the schematics of the measurements.